\documentclass{aastex}
\usepackage{emulateapj5,times,mathptm}
\singlespace
\slugcomment{Accepted for the publication in the {\it Astrophysical Journal 
Letters}}
\def\la{\mathrel{\hbox{\rlap{\hbox{\lower4pt\hbox{$\sim$}}}\hbox{$<$}}}}
\def\ga{\mathrel{\hbox{\rlap{\hbox{\lower4pt\hbox{$\sim$}}}\hbox{$>$}}}}

\shortauthors{Park}
\shorttitle{0103-72.6}

\begin{document}

\title{0103$-$72.6: A New Oxygen-Rich Supernova Remnant 
in the Small Magellanic Cloud}

\author{Sangwook Park\altaffilmark{1}, John P. Hughes\altaffilmark{2}, 
David N. Burrows\altaffilmark{1}, Patrick O. Slane\altaffilmark{3}, 
John A. Nousek\altaffilmark{1}, and Gordon P. Garmire\altaffilmark{1}}

\altaffiltext{1}{Department of Astronomy and Astrophysics, Pennsylvania State
University, 525 Davey Laboratory, University Park, PA. 16802; 
park@astro.psu.edu}
\altaffiltext{2}{Department of Physics and Astronomy, Rutgers University,
136 Frelinghuysen Road, Piscataway, NJ. 08854-8019}
\altaffiltext{3}{Harvard-Smithsonian Center for Astrophysics, 60 Garden Street,
Cambridge, MA. 02138}

\begin{abstract}

0103$-$72.6, the second brightest X-ray supernova remnant (SNR) 
in the Small Magellanic Cloud (SMC), has been observed with the 
{\it Chandra X-Ray Observatory}. Our {\it Chandra} observation 
unambiguously resolves the X-ray emission into a nearly complete, 
remarkably circular shell surrounding bright clumpy emission in 
the center of the remnant. The observed X-ray spectrum for the 
central region is evidently dominated by emission from reverse 
shock-heated metal-rich ejecta. Elemental abundances in this 
ejecta material are particularly enhanced in oxygen and neon, 
while less prominent in the heavier elements Si, S, and Fe. 
We thus propose that 0103$-$72.6 is a new ``oxygen-rich'' SNR,
making it only the second member of the class in the SMC.
The outer shell is the limb-brightened, soft X-ray emission 
from the swept-up SMC interstellar medium. The presence of O-rich 
ejecta and the SNR's location within an H{\small II} region attest 
to a massive star core-collapse origin for 0103$-$72.6. 
The elemental abundance ratios derived from the ejecta suggest an 
$\sim$18~M$_{\odot}$ progenitor star.

\end{abstract}

\keywords {ISM: individual (0103$-$72.6) --- supernova remnants --- 
X-rays: ISM}

\section {\label {sec:intro} INTRODUCTION}

Identifying the explosion type of a supernova remnant (SNR), 
largely between the thermonuclear (Type Ia) and the core-collapse 
(Type II or Ib/Ic) explosions, is difficult, yet important for 
understanding how heavy elements are synthesized and distributed 
into interstellar space. A small number of SNRs are known
to exhibit emission from SN ejecta material which is enhanced in 
O, Ne, and Mg. Based on the nucleosynthesis products of such
explosions, these ``oxygen-rich'' SNRs result from the core-collapse 
of massive progenitor stars. The O-rich SNRs thus provide a useful 
test-ground for the studies of SN nucleosynthesis. The number of 
known O-rich SNRs is small, however: there are three in the Galaxy 
(Cassiopeia A, Puppis A, and G292.0$+$1.8), two (N132D and 
0540$-$69.3) in the Large Magellanic Cloud (LMC), and merely one 
(0102.2$-$7219) in the Small Magellanic Cloud (SMC). Although these 
SNRs are all classified as O-rich SNRs by virtue of the presence of 
ejecta material enriched in O, the nature of the individual objects 
varies significantly from one to another. The detection of more such 
SNRs is thus required for unbiased, systematic investigations of 
this interesting class of SNRs.

0103$-$72.6 is the second brightest X-ray SNR in the SMC 
\citep{seward81,haberl00}. The X-ray images taken with the 
{\it ROSAT}/HRI showed a centrally-peaked morphology and a 
faint outer shell with an $\sim$3$'$ diameter \citep{yoko02}. 
The radio data indicated only a faint shell-like feature roughly
along the X-ray shell \citep{mills82,yoko02}. In the optical band, 
0103$-$72.6 is very faint, probably elliptical, and ring-like with 
nearly the same angular size as the X-ray and the radio remnant 
\citep{math83}. Although the large angular size was suggestive of
a relatively old age ($\sim$10$^4$ yr), the {\it ASCA} data 
suggested elevated metal abundances \citep{yoko02}.
The chemical compositions of the overabundant material were
consistent with a core-collapse Type II SN nucleosynthesis model. 
This result was also supported by the SNR's location within an 
H{\small II} region (DEM S125) (e.g., Filipovi\'c et al. 1998 and 
references therein), which is consistent with a massive progenitor.
 
We report here on the results from our observation of 0103$-$72.6
with the {\it Chandra X-Ray Observatory}. The high resolution
{\it Chandra} data unambiguously resolve the SNR into a 
limb-brightened circular shell surrounding bright, clumpy central 
emission. The central regions are enriched in O and Ne whereas 
the outer shell is dominated by emission from the swept-up SMC 
ISM with low elemental abundances. Based on these results, we 
propose that 0103$-$72.6 is a new O-rich SNR in the SMC.

\section{\label{sec:obs} OBSERVATION \& DATA REDUCTION}

We observed SNR 0103$-$72.6 with the Advanced CCD Imaging
Spectrometer (ACIS) on board the {\it Chandra X-Ray Observatory}
on 2002 August 27 as part of the Guaranteed Time Observation
program. The ACIS-S3 chip, with  very faint (VFAINT) mode, was 
chosen. We screened the raw data in order to utilize the 
VFAINT mode and then filtered by the flight timeline filter. We 
corrected the charge transfer inefficiency (CTI; Townsley 
et al. 2000) with the methods developed by Townsley et al. (2002a).
The data were then screened by the standard filtering with
status and grade. We inspected the overall lightcurve for
any background flares. We found no significant variations of
the lightcurve for the $\sim$3$'$ diameter source region.
After this data reduction, the effective exposure was $\sim$49
ks with $\sim$45000 photons in the 0.3$-$3 keV band. 
The overall spectrum of 0103$-$72.6 is very soft and {\it all}
source photons are detected at $E$ $<$ 3 keV. 

\section{\label{sec:image} X-RAY IMAGES}

\begin{figure}[h]
\figurenum{1}
\centerline{\includegraphics[angle=0,width=3.5in]{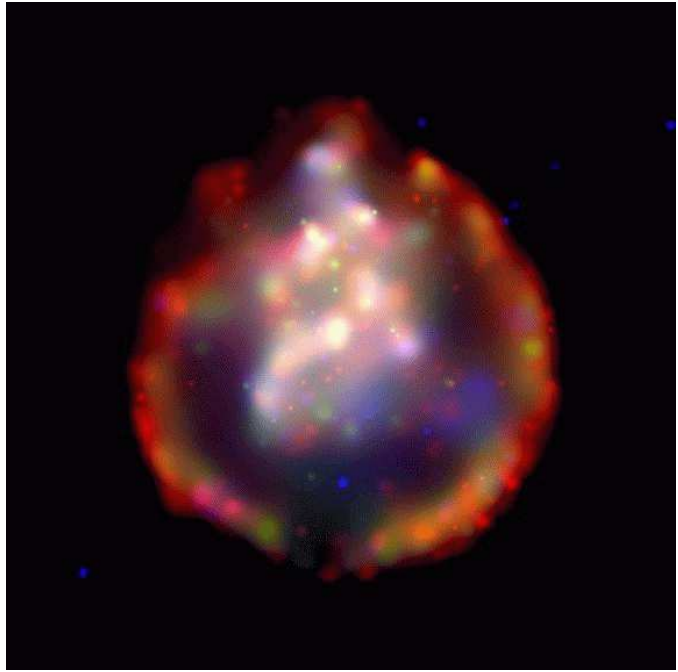}}
\figcaption[]{ The true-color ACIS image of 0103$-$72.6. The red
is 0.3$-$0.75 keV, the green is 0.75$-$0.99 keV, and the blue
is 0.99$-$3.0 keV band images. Each broad-subband image has 
been adaptively smoothed. We note that the bright clumpy features
in the central region are real structures which were resolved
by the ACIS (see Figure~\ref{fig:fig3}). We however caution that faint, 
compact clumps ($\sim$arcsec scale) along the outer shell are likely
artifacts caused by the adaptive smoothing.
\label{fig:fig1}}
\end{figure}

Figure~\ref{fig:fig1} is a ``true-color'' image of SNR 0103$-$72.6.
The high resolution ACIS images clearly resolve the bright central
emission and the limb-brightened outer shell. The outer shell is
remarkably circular and dominated by soft X-ray emission, indicated
in red. The morphological and spectral features of the circumferential 
shell are typical of X-ray emission from ISM swept-up by the 
blast wave shock front. In contrast, the central regions are bright 
in all colors and show a complex mixture of X-ray emitting knots. 
These features are qualitatively consistent with a Sedov picture. 

\begin{figure}[h]
\figurenum{2}
\centerline{\includegraphics[angle=0,width=3.5in]{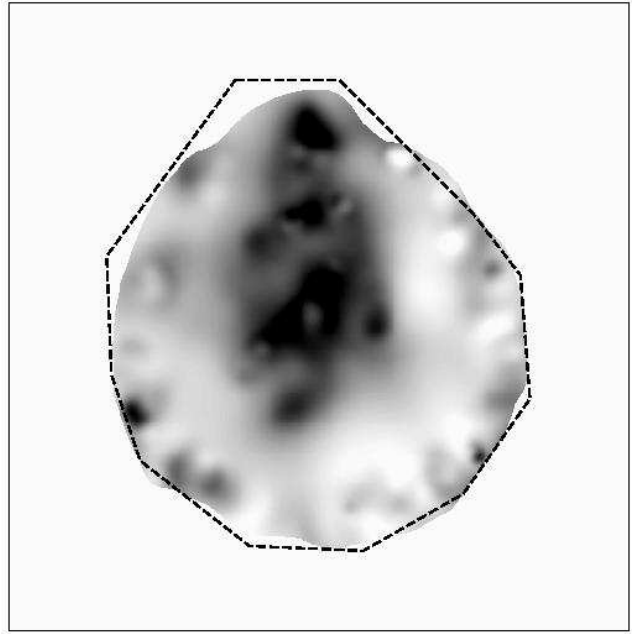}}
\figcaption[]{ The linear gray-scale O EW image of SNR 0103$-$72.6, 
comprised of combined O He$\alpha$ and O Ly$\alpha$ lines. 
The darker gray-scales indicate high EWs: i.e., $\sim$500$-$650 eV
around the central regions, and $\sim$150$-$300 eV otherwise. 
The line emission was extracted from $E$ = 500$-$720 eV. 
The underlying continuum was from $E$ = 300$-$450 eV and $E$ = 
760$-$850 eV, and then logarithmically interpolated to the mean line
energy of 575 eV. Both line and continuum images have been adaptively
smoothed before the calculation of the EW. The EWs were set to zero
near the boundary of the SNR (as marked with a dashed line) in order 
to avoid the noise caused by low background photon statistics. 
\label{fig:fig2}}
\end{figure}

We explored the angular distributions of the X-ray line emission 
by constructing {\it equivalent width} (EW) images for the detected 
elemental species, following the method of Park et al. (2002). 
We generated EW images for O and Ne by selecting photons 
around the broad line features. We present the O EW image in
Figure~\ref{fig:fig2}. The O EW image clearly reveals
enhancements in O line emission in the central regions of 
0103$-$72.6. In contrast, the circumferential shell is fairly 
featureless in the O EW image. These overall features are 
also consistent in the Ne EW image. These EW features support our 
conjecture regarding the overall characteristics of the SNR based 
on the color images: i.e., the central regions are emission from 
metal-rich ejecta heated by the reverse shock, and the outer shell 
is the emission from the forward shock sweeping through the 
ambient ISM.

\section{\label{sec:spec} X-ray Spectra} 

The high resolution ACIS images allow us to perform 
a spatially-resolved spectral analysis of 0103$-$72.6.
For the spectral analysis, we use detector response matrices,
as generated by Townsley et al. (2002b), appropriate for our 
CTI-corrected data. The low energy ($E$ $\la$ 1 keV) quantum 
efficiency (QE) of the ACIS has degraded because of molecular 
contamination on the optical blocking filter. We corrected 
this time-dependent QE degradation by modifying the ancillary 
response function for each extracted spectrum, utilizing 
the IDL ACISABS software\footnote{For the discussion on this 
instrumental issue, see 
http://cxc.harvard.edu/cal/Acis/Cal\_prods/qeDeg/index.html.
The software was developed by George Chartas and is available at
http://www.astro.psu.edu/users/chartas/xcontdir/xcont.html.}.

\begin{figure}[h]
\figurenum{3}
\centerline{\includegraphics[angle=0,width=3.5in]{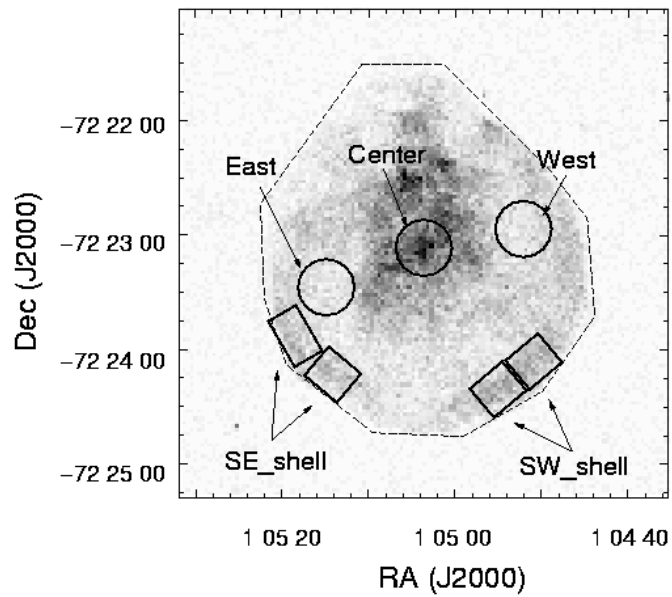}}
\figcaption[]{ The gray-scale broadband image of SNR 0103$-$72.6. 
The image has been rebinned by 4 pixels for the purpose of display. 
Darker gray-scales are higher intensities. Regions used for the 
spectral analysis are marked.
\label{fig:fig3}}
\end{figure}

\begin{figure}[h]
\figurenum{4}
\centerline{\includegraphics[angle=0,width=0.5\textwidth]{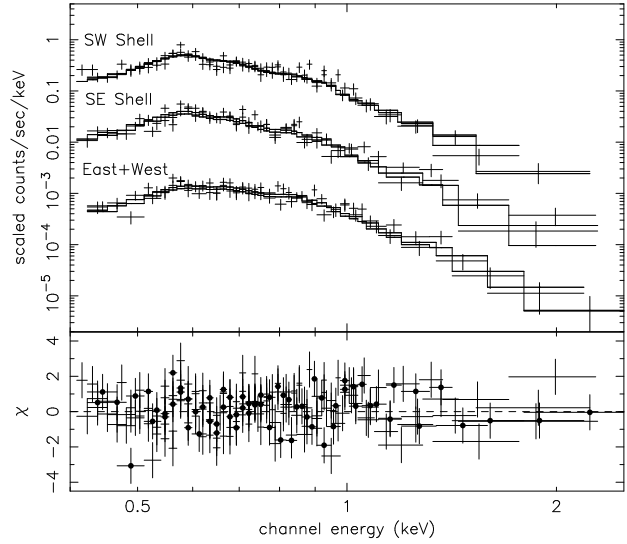}}
\figcaption[]{X-ray spectrum from the swept-up ISM of 0103$-$72.6. 
The individual flux levels have been arbitrarily scaled for the
purpose of display. In the lower panel, residuals from the best-fit 
model for the east+west regions are presented with filled-circles, 
while those for the shell regions are presented as data points without 
filled-circles.
\label{fig:fig4}}
\end{figure}

The presence of the limb-brightened shell, which is remarkably
circular around the centrally-bright emission, indicates a 
significant contribution from the foreground/background swept-up 
ISM to the central excess emission. In order to characterize this 
projected emission from the heated ISM, we first extracted spectra 
from the bright parts of the limb-brightened circumferential 
shell (``SW shell'' and ``SE shell'' regions in 
Figure~\ref{fig:fig3}). This outer shell is most likely 
dominated by emission from the blast wave shock front. 
We thus fitted these spectra simultaneously with a nonequilibrium 
ionization (NEI) plane-parallel shock model \citep{bor01} 
(Figure~\ref{fig:fig4}). Only the normalization was varied 
freely for each individual regional spectrum. These outer shell 
spectra can be best-fitted with $kT$ $\sim$ 0.25$-$0.30~keV, 
depending on the implied foreground column, which is poorly 
constrained ($N_H$ $<$ 1.2 $\times$ 10$^{21}$ cm$^{-2}$). The 
fitted metal abundances are low ($\sim$0.1 solar), which is in 
agreement with typical SMC ISM \citep{russell92}. 

We then extracted spectra from the east and the west side of the 
central excess emission, {\it interior} to the outer shell (i.e., 
``east'' and ``west'' regions in Figure~\ref{fig:fig3}). We 
simultaneously fitted these spectra with an NEI Sedov model 
assuming that the electron temperature immediately behind the shock 
is $kT$~$\sim$~0.25$-$0.3~keV as obtained from the outer shell 
spectrum. Only the normalizations were allowed to vary independently. 
While the Sedov model corresponds to the full temperature and density 
distribution of the swept-up component of an entire SNR, we use this 
model as a first-order approximation to account for the expected 
temperature distribution for the plasma in the SNR's interior. The 
goal of these fits is to describe the projected blast wave emission 
seen in the central region in combination with the ejecta component.
These spectra can be described by $kT$~$\sim$~0.26~keV 
and low SMC-like abundances of 0.12$^{+0.08}_{-0.04}$ solar 
(Figure~\ref{fig:fig4}; Table~\ref{tbl:tab1}). 

\begin{figure}[h]
\figurenum{5}
\centerline{\includegraphics[angle=0,width=0.5\textwidth]{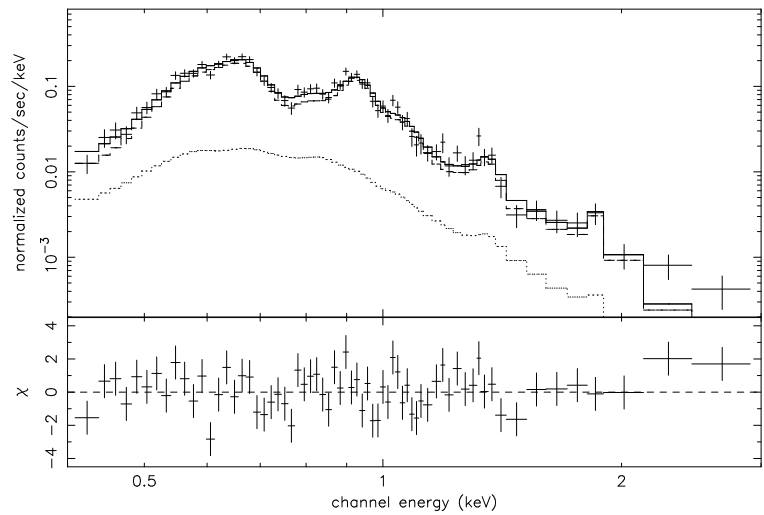}}
\figcaption[]{X-ray Spectrum of the O-rich ejecta as extracted from
the ``center'' region. The projected emission from the swept-up ISM
has been included with a Sedov component (the dotted line).
\label{fig:fig5}}
\end{figure}

The central region spectrum has been extracted from the ``center'' 
region (Figure~\ref{fig:fig3}). We fitted this spectrum with a
two-component model: i.e., a Sedov component for the projected
emission from the swept-up ISM, and a plane-shock model for the
central ejecta component (Figure~\ref{fig:fig5}). We fixed the 
Sedov component parameters at the best-fit values obtained above. 
This Sedov component for the projected emission from the swept-up 
ISM comprises $\sim$15\% of the observed flux for 
the ``center'' region. The best-fit ejecta electron temperature 
($kT$~$\sim$~0.53~keV) is higher than that of the blast wave shock 
front as measured from the outer shell (Table~\ref{tbl:tab1}). 
The elemental abundances for the ejecta component were varied freely 
for O, Ne, Mg, Si and Fe while others are fixed at the SMC values 
\citep{russell92}. The fitted abundances are also higher than those 
of the SMC: particularly, O and Ne abundances are 4$-$6 times higher 
than typical SMC abundances. Mg and Si are moderately elevated 
($\sim$1.5 times the SMC), and Fe abundance ($\sim$0.3 times the SMC) 
is very low. The S abundance is unconstrained with the current
data and thus fixed at the SMC abundance. The improvement made by 
fitting the S abundance is statistically negligible.

\section{\label{sec:disc} DISCUSSION}

The X-ray spectrum from the bright central region of 0103$-$72.6 
indicates the presence of ejecta material which is particularly
enriched in O and Ne. We thus propose that 0103$-$72.6 is a new
O-rich SNR, based on the results of our X-ray spectral 
analysis. 0103$-$72.6 is then the second member of this class
in the SMC in addition to SNR 0102.2$-$7219 \citep{dopita84}. 
The O- and Ne-dominated nature of the ejecta of 0103$-$72.6,
with the relative absence of high-Z species, particularly Fe, is 
similar to that of 0102.2$-$7219 \citep{blair00,sasaki01} and the 
Galactic O-rich SNR G292.0+1.8 \citep{park03a}. This abundance 
distribution however differs from that of Cas A, which shows 
significant Si-, S-, and Fe-rich ejecta \citep{hughes00a}. 
The physical location within an H{\small II} region and the
detected O-rich nature are typical signatures of a core-collapse
origin for 0103$-$72.6. We thus compare the measured elemental
abundances of 0103$-$72.6 with a Type II SN nucleosynthesis model
(Table~\ref{tbl:tab2}). Although the measured elemental abundances
have systematic uncertainties depending on the embedded contribution 
from the background swept-up ISM component, the quoted abundance 
ratios are nearly independent of these uncertainties. The 
measured abundance ratios are in plausible agreement with a Type 
II nucleosynthesis model for an 18~M$_{\odot}$ progenitor star. 

Assuming that 0103$-$72.6 is in the Sedov phase, we may derive 
the SNR parameters by taking advantage of the spatial information 
obtained by the high-resolution ACIS images.
Based on the best-fit volume emission measure 
($EM$~=~5.4~$\times$~10$^{58}$~cm$^{-3}$ by combining four fitted 
shell regions; Table~\ref{tbl:tab1}), we derive the postshock 
electron density, $n_e$~=~0.79 cm$^{-3}$ for the blast wave shock 
front. In this estimation, we have assumed that the average outer 
radius of the SNR is $R$ $\sim$ 85$^{\prime\prime}$ (corresponding
to $\sim$24.7 pc at $d$ = 60 kpc) and that the blast wave shell 
is $\sim$20$^{\prime\prime}$ thick, based on the ACIS image.
The average depth through our shell regions is then $\sim$23 pc which 
implies a total volume for the four shell regions of $\sim$1 $\times$ 
10$^{59}$ cm$^3$. The pre-shock hydrogen density is
then $n_0$ $\sim$ 0.16 cm$^{-3}$ assuming a factor of 4 density
compression behind shock and a mean charge state of $n_e$ = 1.2~$n_H$. 
Assuming uniform density for the ambient ISM, we estimate the mass 
swept-up by the blast wave to be $M_{swept}$~$\sim$~570 M$_{\odot}$.
The best-fit electron temperature of the shock front ($kT$ = 0.3 keV) 
corresponds to a shock velocity of $v$ $\sim$ 480 km s$^{-1}$, 
under the assumption of electron-ion temperature equilibration.
We then derive the SNR age of $t$ $\sim$ 1.8 $\times$ 10$^4$ yr
and the SN explosion energy of $E_0$ $\sim$ 1.2 $\times$ 10$^{51}$
ergs based on the Sedov solution. 0103$-$72.6 thus appears to be
the oldest known example of an O-rich SNR.
The detection of significant amounts of ejecta material in 
0103$-$72.6 adds yet another such an example of the growing number 
of middle-aged SNRs with observable ejecta (e.g., Park et al. 2003b 
and references therein). 

We make a brief morphological comparison between the two SMC O-rich
SNRs, 0103$-$72.6 and 0102.2$-$7219. 0102.2$-$7219 is a young 
SNR ($\sim$10$^3$ yr) with a small angular size of 
$\sim$44$^{\prime\prime}$ diameter, and the X-ray morphology shows
a bright ring-like structure embedded in a relatively faint 
diffuse emission extending beyond the bright ring \citep{gaetz00}. 
This morphology was interpreted as the reverse shock front heating 
the central O-rich ejecta material (i.e., the sharp inner boundary 
of the bright ring) and the blast wave shock front propagating into 
the ambient ISM (i.e., the outer boundary of the faint extended 
diffuse emission) \citep{gaetz00}. In comparison, as we have just 
determined, 0103$-$72.6 has a larger angular size (190$^{\prime\prime}$) 
and is significantly older (1.8 $\times$ 10$^4$ yr). The reverse 
shock-heated ejecta material shows a relatively well-defined, but 
complex structure near the center of the SNR. The blast wave shock 
front is represented by the limb-brightened, soft shell-like feature 
surrounding the central ejecta. It appears that the reverse shock
has reached the center of the SNR in 0103$-$72.6, while it has
reached only $\sim$60\% of the current blast wave shock radius
in 0102.2$-$7219 \citep{gaetz00}. The blast wave shock front
of 0103$-$72.6 has swept up significant amounts of ambient ISM and 
slowed down ($v$~$\sim$~480~km~s$^{-1}$) to produce a 
limb-brightened soft X-ray shell, whereas 0102.2$-$7219 shows a 
high velocity ($v$~$\sim$~6000~km~s$^{-1}$; Hughes et al. 2000b)
shock front without significant limb-brigtening 
and presumably less swept-up ISM. Both of the SNRs have a remarkably 
circular outer boundary, and thus likely uniform surrounding ISM 
distributions. The morphological characteristics of these two SMC 
SNRs then appear to be primarily caused by the substantial difference 
in the ages ($\sim$10$^{3}$ yr vs. $\sim$10$^4$ yr): i.e., 
0103$-$72.6 appears to be a more evolved version of 0102.2$-$7219.

\acknowledgments
This work has been supported in parts by NASA contract NAS8-01128. 
JPH was supported by {\it Chandra} grants GO2-3069X and GO3-4086X. 
POS was supported by NASA contract NAS8-39073.

\begin{deluxetable}{lcc}
\footnotesize
\tablecaption{Spectral parameters of 0103$-$72.6.
\label{tbl:tab1}}
\tablewidth{0pt}
\tablehead{ & \colhead{ISM\tablenotemark{a}} & 
\colhead{Ejecta\tablenotemark{b}} }
\startdata
$N_H$ (10$^{21}$ cm$^{-2}$) & $<$ 1.2 & 1.1$^{+0.6}_{-0.4}$ \\
$kT$ (keV) & 0.30$^{+0.12}_{-0.05}$ & 0.53$^{+0.25}_{-0.18}$ \\
$n_et$ (10$^{11}$ cm$^{-3}$ s) & 1.6$^{+1.2}_{-0.7}$ & 1.4$^{+0.3}_{-0.3}$ \\
$EM$ (10$^{58}$ cm$^{-3}$) & 5.4$^{+1.1}_{-1.1}$ & 0.8$^{+0.3}_{-0.3}$ \\
$\chi$$^{2}$/$\nu$ & 169.0/137 & 86.7/58 \\
\enddata
\tablenotetext{a}{The best-fit plane-shock model from the combined four
shell regions.}
\tablenotetext{b}{The plane-shock component of the center region.}
\end{deluxetable}
 
\begin{deluxetable}{cccccccc}
\footnotesize
\tablecaption{Comparisons of the Abundance Ratios.
\label{tbl:tab2}}
\tablewidth{0pt}
\tablehead{ \colhead{Abundance} & \colhead{Center} & 
\colhead{SMC\tablenotemark{a}} & & & \colhead{Model\tablenotemark{b}} & & \\
\colhead{Ratio} & \colhead{0103$-$72.6} & & \colhead{15 M$_{\odot}$} & 
\colhead{18 M$_{\odot}$} & \colhead{20 M$_{\odot}$} & 
\colhead{25 M$_{\odot}$} & \colhead{40 M$_{\odot}$} }
\startdata
O/Ne & 4.1$^{+1.2}_{-1.1}$ & 5.8 & 13.0  & 5.3 & 7.1 & 5.9 & 16.0 \\ 
O/Mg & 31.5$^{+12.5}_{-14.7}$ & 11.2 & 15 & 24 & 12 & 19 & 26 \\
O/Si & 27.1$^{+12.8}_{-14.9}$ & 10.0 & 7.9 & 15 & 25 & 45 & 136 \\
\enddata
\tablenotetext{a}{Russell \& Dopita (1992).}
\tablenotetext{b}{Type II SN nucleosynthesis model by Nomoto et al. (1997).}
\end{deluxetable}

\end{document}